\documentclass{llncs}
\begin{document}
%
%\frontmatter          % for the preliminaries
%
%\pagestyle{headings}  % switches on printing of running heads
\pagestyle{empty}
\bibliographystyle{splncs}
\addtocmark{Astronomical Pipelines} % additional mark in the TOC
\title{Using Fuzzy Logic for Automatic Analysis of Astronomical Pipelines}
\titlerunning{Astronomical Pipelines}
\author{Lior Shamir \and Robert J. Nemiroff}
\authorrunning{Lior Shamir \& Robert J. Nemiroff}
\tocauthor{Lior Shamir and Robert J. Nemiroff (Michigan Technological University)}
\institute{Michigan Technological University, Houghton MI 49931, USA \\
\email{lshamir@mtu.edu}}

\maketitle

\begin{abstract}
Fundamental astronomical questions on the composition of the universe, the abundance of Earth-like planets, and the cause of the brightest explosions in the universe are being attacked by robotic telescopes costing billions of dollars and returning vast pipelines of data.  The success of these programs depends on the accuracy of automated real time processing of the astronomical images.  In this paper the needs of modern astronomical pipelines are discussed in the light of fuzzy-logic based decision-making.  Several specific fuzzy-logic algorithms have been develop for the first time for astronomical purposes, and tested with excellent results on data from the existing Night Sky Live sky survey.  \\
\end{abstract}

\section{Introduction}
In the past few years, pipelines providing astronomical data have been becoming increasingly important. The wide use of robotic telescopes has provided significant discoveries, and sky survey projects are now considered among the premier projects in the field astronomy. In this paper we will concentrate on the ground based missions, although future space based missions like Kepler, SNAP and JWST will also create significant pipelines of astronomical data.

Pan-STARRS \cite{Kai04}, a 60 million dollar venture, is being built today and completion is expected by 2006. Pan-STARRS will be composed of 4 large telescopes pointing simultaneously at the same region of the sky. With coverage of 6000 degrees$^{2}$ per night, Pan-STARRS will look for transients that include supernovas, planetary eclipses, and asteroids that might pose a future threat to Earth. Similarly but on a larger scale, ground-based LSST \cite{Tys02} is planned to use a powerful 8.4 meter robotic telescope that will cover the entire sky every 10 days. LSST will cost \$200M, be completed by 2012, and produce 13 terabytes per night. In addition, many smaller scale robotic telescopes are being deployed and their number is growing rapidly.  

However, in the modern age of increasing bandwidth, human identifications are many times impracticably slow. Therefore, efforts toward the automation of the analysis of astronomical pipelines have been gradually increasing. In this paper we present fuzzy logic based algorithms for two basic problems in astronomical pipeline processing which are rejecting cosmic ray hits and converting celestial coordinates to image coordinates.

\section {Fuzzy Logic Based Coordinate Transformations}

Useful automatic pipeline processing of astronomical images depends on accurate algorithmic decision making.  For previously identified objects, one of the first steps in computer-based analysis of astronomical pictures is an association of each object with a known catalog entry.  This necessary step enables such science as automatically detected transients and automated photometry of stars. Since computing the topocentric coordinates of a given known star at a given time is a simple task, transforming the celestial topocentric coordinates to image $(x,y)$ coordinates might provide the expected location of any star in the frame. However, in an imperfect world of non-linear wide-angle optics, imperfect optics, inaccurately pointed telescopes, and defect-ridden cameras, accurate transformation of celestial coordinates to image coordinates is not always a trivial first step.

On a CCD image, pixel locations can be specified in either Cartesian or polar coordinates.  Let $x_{\mathrm zen}$ be the $x$ coordinate (in pixels) of the zenith in the image, and $y_{\mathrm zen}$ be the $y$ coordinate of the zenith.  
In order to use polar coordinates, it is necessary to transform the topocentric celestial coordinates (Azimuth,Altitude) to a polar distance and angle from ($x_{\mathrm zen},y_{\mathrm zen}$).

\subsection{The Fuzzy Logic Model}

The fuzzy logic model is built based on manually identified reference stars. Each identified star contributes an azimuth and altitude (by basic astronomy) and also an angle and radial distance (by measurement from the image). These provide the raw data for constructing a mapping between the two using the fuzzy logic model. In order to transform celestial coordinates into image coordinates, we need to transform the azimuth and altitude of a given location to polar angle in the image from ($x_{\mathrm zen},y_{\mathrm zen}$) and the radial distance (in pixels) from ($x_{\mathrm zen},y_{\mathrm zen}$).

In order to obtain the coordinates transformation, we build a fuzzy logic model based on the reference stars. The model has two antecedent variables which are {\it altitude} and {\it azimuth}. The {\it azimuth} is fuzzified using pre-defined four fuzzy sets {\it North},{\it East}, {\it South} and {\it West}, and each fuzzy set is associated with a Gaussian membership function. The {\it altitude} is fuzzified using fuzzy sets added to the model by the reference stars such that each fuzzy set is associated with a triangular membership function that reaches its maximum of unity at the reference value and intersects with the x-axis at the points of maximum of its neighboring reference stars.

The fuzzy rules are defined such that the antecedent part of each rule has two fuzzy sets (one for altitude and one for azimuth) and the consequent part has one crisp value which is the distance (in pixels) from ($x_{\mathrm zen},y_{\mathrm zen}$). The reasoning procedure is based on {\it product} inferencing and {\it weighted average} defuzzification, which is an efficient defuzzification method when the fuzzy logic model is built according to a set of singleton values \cite{Tak85}.

\subsection{Application to the {\it Night Sky Live} Sky Survey}

The algorithm has been implemented for the {\it Night Sky Live!} \cite{Nem99} project, which deploys 10 all-sky CCD cameras called {\it CONCAM} at some of the world`s premier observatories covering almost the entire night sky. The pictures are 1024 $\times$ 1024 FITS images, which is a standard format in astronomical imaging.

The algorithm allows practically 100 percent chance of accurate identification for NSL stars down to a magnitude of 5.6. This accurate identification allows systematic and continuous monitoring of bright star, and photometry measurements are constantly being recorded and stored in a database. The automatic association of PSFs to stars is also used for optical transient detection.
 
\section{Cosmic Ray Hit Rejection Using Fuzzy Logic}

The presence of cosmic ray hits in astronomical CCD frames is frequently considered as a disturbing effect. Except from their annoying presence, cosmic ray hits might be mistakenly detected as true astronomical sources. Algorithms that analyze astronomical frames must ignore the peaks caused by cosmic ray hits, yet without rejecting the peaks of the true astronomical sources.

\subsection{A Human Perception-Based Fuzzy Logic Model}
\label{manual_detection}

Cosmic ray hits in astronomical exposures are usually noticeably different then point spread functions of true astronomical sources. Cosmic ray hits are usually smaller than true PSFs, and their edges are usually sharper. An observer trying to manually detect cosmic ray hits would probably examine the edges and the surface size each peak. Since some of the cosmic ray hits have only one or two sharp edges, it is also necessary to examine the sharpest edge of the PSF. For instance, if the surface size of the peak is very small, it has sharp edges and the sharpest edge is extremely sharp, it would be classified as a cosmic ray hit. If the surface size of the peak is large and its edges are not sharp, it would be probably classified as a PSF of an astronomical source.

In order to model the intuition described above, we defined 3 antecedent fuzzy variables: the surface size of the PSF, the sharpness of the sharpest edge and the average sharpness of the edges. The consequent variable is the classification of the peak, and its domain is \{1,0\}. Since astronomical images typically contain 1 to 16 million pixels, the triangular membership functions are used for their low computational cost.

The fuzzy rules are defined using the membership functions of the antecedent variables and the domain of the consequent variable (\{0,1\}), and are based on the natural language rules of intuition. For instance, the rules of intuition described earlier in this section would be compiled into the fuzzy rules:\newline
{\it small} $\wedge$ {\it sharp} $\wedge$ {\it extreme} $\longmapsto$ 1 \newline
{\it large} $\wedge$ {\it low} $\wedge$ {\it low}  $\longmapsto$ 0 \newline
The computation process is based on {\it product} inferencing and {\it weighted average} defuzzification \cite{Tak85}, and the value of the consequent variable is handled such that value greater than $0.5$ is classified as a cosmic ray hits. Otherwise, the value is classified as a non-cosmic ray hits.

\subsection{Using the Fuzzy Logic Model}
\label{using_the_model}

The fuzzy logic model is used in order to classify peaks in the frame as cosmic ray hits or non-cosmic ray hits. In the presented method, searching for peaks in a FITS frame is performed by comparing the value of each pixel with the values of its 8 neighboring pixels. If the pixel is equal or brighter than {\it all} its neighboring pixels, it is considered as a center of a peak. After finding the peaks in the frame, the fuzzy logic model is applied to the peaks in order to classify them as cosmic ray hits or non-cosmic ray hits.

Measurements of the performance of the algorithm were taken using 24 {\it Night Sky Live} exposures. Each NSL frame contains an average of 6 noticeable cosmic ray hits brighter than 20$\sigma$, and around 1400 astronomical sources. Out of 158 cosmic ray hits that were tested, the algorithm did not reject 4, and mistakenly rejected 6 true astronomical sources out of a total of 31,251 PSFs. These numbers are favorably comparable to previously reported cosmic ray hit rejection algorithms, and the presented algorithm also has a clear advantage in terms of computational complexity.

\section{Conclusion}

The emerging field of robotic telescopes and autonomous sky surveys introduces a wide range of problems that require complex decision making. We presented solutions to two basic problems, which are star recognition and cosmic ray hit rejection. We showed that fuzzy logic modeling provides the infrastructure for complex decision making required for automatic analysis of astronomical frames, yet complies with the practical algorithmic complexity constraints introduced by the huge amounts of data generated by the astronomical pipelines.

\end{document}